\documentstyle[12pt]{article}

\begin{document}

\begin{flushright}
IMSc/2011/11/15 
\end{flushright} 

\vspace{2mm}

\vspace{2ex}

\begin{center}
{\large \bf Static brane--like vacuum solutions } \\ 

\vspace{2ex}

{\large \bf in $D \ge 5 \;$ dimensional spacetime } \\

\vspace{2ex}

{\large \bf with positive ADM mass but no horizon } \\

\vspace{8ex}

{\large  S. Kalyana Rama}

\vspace{3ex}

Institute of Mathematical Sciences, C. I. T. Campus, 

Tharamani, CHENNAI 600 113, India. 

\vspace{1ex}

email: krama@imsc.res.in \\ 

\end{center}

\vspace{6ex}

\centerline{ABSTRACT}

\begin{quote} 

We describe static, brane--like, solutions to vacuum Einstein's
equations in $D = n + m + 2 \;$ dimensional spacetime with $m
\ge 2$ and $n \ge 1 \;$. These solutions have positive ADM mass
but no horizon. The curvature invariants are finite everywhere
except at $r = 0 \;$ where $r$ is the radial coordinate in the
$m + 1$ dimensional space. The presence of $n \ge 1$ extra
dimensions is crucial for these properties. Such solutions may
be naturally anticipated if Mathur's fuzzball proposal for black
holes is correct.

\end{quote}

\vspace{2ex}


\newpage

\centerline{\bf 1. Introduction}  

\vspace{4ex}

We consider a $D = n + m + 2$ dimensional spacetime with $m \ge
2 \;$. The $n$ dimensional space may be taken to be compact and
toroidal, or to be $\mathbf{R^n} \;$. We assume a static,
brane--like ansatz for the line element, hence the metric
components are independent of time and the $n$ dimensional
coordinates, and depend only on the radial coordinate $r$ of the
$m + 1$ dimensional space. 

We study the solutions to vacuum Einstein's equations ${\cal
R}_{M N} = 0 \;$ where ${\cal R}_{M N}$ is the $D$ dimensional
Ricci tensor. The solutions are given by 
\[
d s^2 = - e^{2 a_0 F} d t^2 + \sum_i e^{2 a_i F} (d x^i)^2
+ \frac{d r^2}{f} + r^2 d \Omega_m^2 
\]
where $i = 1, 2, \cdots, n \;$ and $(a_0, a_i)$ are constants
obeying $a_0 + \sum_i a_i = \frac{1}{2} \;$. The solutions are
required to be asymptotically Minkowskian with positive ADM
mass. Einstein's equations can be solved in a closed form for
$F(f) \;$ and $r(f) \;$ which, however, is cumbersome to
analyse. But it turns out that the qualitative properties of the
evolution of $f(r)$ and $e^{F(r)}$ can be understood without
this closed form. We find the following results. 

\vspace{2ex}

For $K = a_0^2 + \sum_i a_i^2 - \frac{1}
{4} = 0 \;$, we get $e^F = f = 1 - \frac{M_\infty} {r^{m - 1}}
\;$ where $M_\infty > 0 \;$ is a constant proportional to ADM
mass. Standard black $n-$brane solution then follows for $2 a_0
- 1 = a_i = 0 \;$; for other values of $(a_0, a_i)$ satisfying
$K = 0 \;$, one obtains solutions studied {\em e.g.} in
\cite{k05}.

\vspace{2ex}

For $K > 0 \;$, the solutions have novel properties. For these
solutions, as $r$ decreases from $\infty$ to $0 \;$ :

\begin{itemize}

\item 

$ \; f(r)$ decreases from $1 \;$, reaches a minimum $f_0 > 0 \;$
at some $r_0 \;$, increases to $1$ again at some $r = r_1 < r_0
\;$, and then increases to $\infty$ in the limit $r \to 0
\;$. In particular, $f(r)$ always remains positive and non
zero. 

\item 

$e^{F(r)}$ decreases monotonically from $1$ to $0 \;$. 

\item 

Define a `mass' function $M(r) = r^{m - 1} \; (1 - f) \;$. As
$r$ decreases from $\infty$ to $0 \;$, $\; M(r)$ decreases from
$M_\infty$ but remains positive for $r_1 < r \;$, vanishes at
$r_1 \;$, becomes negative for $r < r_1 \;$, and decreases to $-
\infty$ in the limit $r \to 0 \;$. The evolution of $M(r)$ from
$M_\infty$ to $- \infty$ is monotonic.

\vspace{2ex}

These features are in contrast to the standard Schwarzschild or
black $n-$brane solution where, as $r$ decreases from $\infty$
to $0 \;$, $M(r) = M_\infty \;$ remains constant and $e^F = f =
1 - \frac {M_\infty} {r^{m - 1}} \;$ decreases from $1 \;$,
vanishes at some $r_h \;$, and decreases to $- \infty$ in the
limit $r \to 0 \;$.

\end{itemize}

Note that $f$ and $e^F$ and, hence, all metric components remain
non zero and finite for $0 < r \le \infty \;$. This implies that
there is no horizon \footnote{ Solutions with negative masses in
the interior and with no horizon occur in \cite{zurek, thooft}
which study the back reaction of Hawking radiation in four
dimensional spacetime. Solutions with a central singularity and
with no horizon occur in \cite{date1, date2} which study static
solutions with incoming radiation matching the outgoing one; the
solutions in \cite{date1} can be matched onto negative mass
Schwarzschild solutions.} and that the curvature invariants are
all finite for $0 < r \le \infty \;$. As $r \to 0 \;$, we have
$f \to \infty \;$ and $e^F \to 0 \;$. In this limit, the tidal
forces also diverge and there is a curvature singularity.

The presence and the role of the $n-$dimensional space is
crucial for these properties of the solutions. The absence of
the $n-$dimensional space, or the trivialty of its metric, means
that $a_i = K = 0 \;$ which leads to the standard Schwarzschild,
or black $n-$brane, solution. We assume that $a_i \ne 0 \;$
generically and, further, that $K > 0 \;$ which then lead to the
solutions described here.

We discuss how the $K > 0$ solutions may be naturally
anticipated if one assumes that Mathur's fuzzball proposal
\cite{f1} -- \cite{f5} for black holes is correct.

This paper is organised as follows. In section {\bf 2}, we
present the equations and write them in a convenient form. In
section {\bf 3}, we analyse the equations and describe the
evolution of $f$, $\; e^F$, and $M \;$. In section {\bf 4}, we
discuss the physical relevance of the $K > 0$ solutions, and
conclude in Section {\bf 5}.


\vspace{4ex}

\centerline{\bf 2. Einstein's equations in vacuum}

\vspace{2ex}

Consider a $D = n + m + 2$ dimensional spacetime with $m \ge 2
\;$. Consider static, brane--like ansatz for the line element $d
s$ given by
\begin{equation}\label{ds}
d s^2 = - e^{2 \psi} d t^2 + \sum_i e^{2 \lambda^i} (d x^i)^2
+ e^{2 \lambda} d r^2 + e^{2 \sigma} d \Omega_m^2
\end{equation}
where $d \Omega_m$ is the standard line element on an $m$
dimensional unit sphere, $i = 1, 2, \cdots, n \;$, and $(\psi,
\lambda^i, \lambda, \sigma)$ are all functions of $r$ only.
Such an ansatz is suitable for describing intersecting brane
configurations of string/M theory. We take the $n$ dimensional
space described by $x^i$ coordinates to be compact and toroidal
but it may also be taken to be $\mathbf{R^n} \;$. 

Let $\Lambda = \psi + m \sigma + \sum_i \lambda^i \;$. The
vacuum Einstein's equations ${\cal R}_{M N} = 0 \;$, where
${\cal R}_{M N}$ is the $D$ dimensional Ricci tensor, then give
\begin{eqnarray}
\Lambda_r^2 - ( \psi_r^2 + m \; \sigma_r^2 
+ \sum_i (\lambda_r^i)^2 ) & = & 
m \; (m - 1) \; e^{2 \lambda - 2 \sigma} \label{rr} \\
\psi_{r r} + (\Lambda_r - \lambda_r) \; \psi_r & = & 0
\label{00} \\
\lambda_{r r}^i + (\Lambda_r - \lambda_r) \; \lambda_r^i & = & 0
\label{ii} \\
\sigma_{r r} + (\Lambda_r - \lambda_r) \; \sigma_r & = & 
(m - 1) \; e^{2 \lambda - 2 \sigma} \label{aa} 
\end{eqnarray}
where $r-$subscripts denote derivatives with respect to $r \;$.
Equations (\ref{00}) and (\ref{ii}) imply that
\begin{equation}\label{aai}
\psi = a_0 \; F 
\; \; , \; \; \; 
\lambda^i = a_i \; F 
\end{equation}
where $a_0$ and $a_i$ are constants and the function $F(r)$ is
defined by
\begin{equation}\label{Fdefn}
F_{r r} + (\Lambda_r - \lambda_r) \; F_r = 0 
\; \; \;  \Longrightarrow \; \; \; 
e^{\Lambda - \lambda} \; F_r = (m - 1) \; {\cal N} \; \; ,
\end{equation}
with ${\cal N}$ an integration constant. Choose $e^\sigma = r
\;$ so that $r$ denotes the physical size of the $m$ sphere.
Then
\begin{equation}\label{A}
\Lambda = m \; ln \; r + A \; F 
\; \; , \; \; \; A = a_0 + \sum_i a_i \; \; . 
\end{equation}
Further, replace $\lambda(r)$ by an equivalent function $f(r)$
and also define a `mass' function $M(r) \;$ analogous to that in
the study of stars, as follows:
\begin{equation}\label{fM}
e^{- 2 \lambda} = f(r) = 1 - \frac{M(r)}{r^{m - 1}} \; \; .
\end{equation}
The $D$ dimensional line element $d s$ is now given by 
\begin{equation}\label{ds1}
d s^2 = - e^{2 a_0 F} d t^2 + \sum_i e^{2 a_i F} (d x^i)^2
+ \frac{d r^2}{f} + r^2 d \Omega_m^2 \; \; .
\end{equation}
Equations (\ref{rr}), (\ref{aa}), and (\ref{Fdefn}) now give,
after some rearrangements, 
\begin{eqnarray}
2 A \;
f \; (r F_r) & = & (m - 1) \; (1 - f) 
+ \frac{K}{m} \; f \; (r F_r)^2 \label{rr1} \\
2 A \;
f \; (r F_r) & = & 2 (m - 1) \; (1 - f) - r f_r 
\label{aa1} \\
e^{2 A \; F} \; f \; (r F_r)^2 & = & 
\frac{(m - 1)^2 \; {\cal N}^2}{r^{2 (m - 1)}} 
\label{Fdefn1}
\end{eqnarray}
where $K = a_0^2 + \sum_i a_i^2 - A^2 \;$. From equations
(\ref{rr1}) and (\ref{aa1}), we have
\begin{equation}\label{rr-aa}
\frac{K}{m} \; f \; (r F_r)^2 = (m - 1) \; (1 - f) - r f_r 
\; \; .
\end{equation}
We set $2 A = 1 \;$ with no loss of generality since this just
amounts to defining $F$ by $F = 2 \; (\psi + \sum_i \lambda^i)
\;$. Thus, we have \footnote{More generally, one may also
consider $\sigma = ln \; r + c \; F \;$ and $e^{- 2 (\lambda - c
F)} = f(r) \;$. Then $A$, $K$, and the condition $2 A = 1$ are
replaced by $A = a_0 + m c + \sum_i a_i \;$, $\; K = a_0^2 + m
c^2 + \sum_i a_i^2 - A^2 \;$, and $\; 2 (A - c) = 1 \;$.}
\begin{equation}\label{AK}
A = a_0 + \sum_i a_i = \frac{1}{2} \; \; , \; \; \; 
K = a_0^2 + \sum_i a_i^2 - \frac{1}{4} \; \; . 
\end{equation}
Note that if $a_i$ do not all vanish then, generically, $K \ne 0
\;$. \footnote{ Indeed, we have $- \; \frac{n}{4 (n + 1)}
\le K \le \infty \;$ which can be derived as follows. Let
$\vec{a} = (a_0, a_1, \cdots, a_n)$ and $\vec{\mathbf{1}} = (1,
1, \cdots, 1)$ be two $(n + 1)-$component vectors. Then $A =
\vec{a} \cdot \vec{\mathbf{1}} \;$, $\; A^2 = (n + 1) \vert
\vec{a} \vert^2 cos^2 \theta \;$, and $K = \vert \vec{a} \vert^2
- A^2 \;$. The inequality follows since $\vert \vec{a} \vert^2 =
\frac{A^2} {(n + 1) \; cos^2 \theta} \ge \frac{ A^2}{(n + 1)}
\;$ and $A = \frac{1}{2} \;$.} Now using equation (\ref{aa1})
for $(r F_r) \;$ in equations (\ref{rr-aa}) and (\ref{Fdefn1}),
we have
\begin{equation}\label{freqn}
\frac{K}{m} \; \left( 2 (m - 1) \; (1 - f) - r f_r \right)^2 
= f \; \left( (m - 1) \; (1 - f) - r f_r \right) \; \; , 
\end{equation}
which is to be solved for $f(r) \;$, and
\begin{equation}\label{F(r)}
e^F = \frac{(m - 1)^2 \; {\cal N}^2 \; f}{r^{2 (m - 1)} \; 
\left( 2 (m - 1) \; (1 - f) - r f_r \right)^2} \; \; . 
\end{equation}
Thus, once $f(r)$ is known, $\; e^F$ and the line element $d s$
are completely determined. In order to obtain asymptotically
Minkowskian solutions with positive ADM mass, we require that,
in the limit $r \to \infty \;$,
\begin{equation}\label{bcr}
e^F \to 1 
\; \; \; , \; \; \; \; 
f(r) \to 1 - \frac{M_\infty}{r^{m - 1}} 
\; \; \; , \; \; \; \; M_\infty = const > 0 \; \; . 
\end{equation}

Note that the condition on $e^F$ implies that ${\cal N}^2 =
M_\infty^2 \;$, and that the condition on $f$ implies that $r
f_r \to (m - 1) (1 - f) \;$ irrespective of whether $M_\infty
\;$ is positive or negative. Also, in the limit $r \to \infty
\;$, it follows from equations that (\ref{rr1}) --
(\ref{Fdefn1})
\begin{eqnarray}
f & = & 1 - \frac{M_\infty}{r^{m - 1}} 
+ \frac{(m - 1) \; K}{m} \; \left( \frac{M_\infty}{r^{m - 1}}
\right)^2
+ \frac{(m - 1) \; K}{2 m} \; \left( \frac{M_\infty}{r^{m - 1}}
\right)^3 + \cdots
\nonumber \\ 
e^F & = & 1 - \frac{M_\infty}{r^{m - 1}} 
+ \frac{(m - 1) \; K}{6 m} \; \left( \frac{M_\infty}{r^{m - 1}}
\right)^3 + \cdots 
\label{asympeF} 
\end{eqnarray}
where $\cdots$ denote terms of ${\cal O} \left( r^{- 4 (m - 1)}
\right) \;$. ADM mass is given, using the asymptotic form of
$e^F \;$, by
\begin{equation}\label{adm}
M_{ADM} = \frac{m \; \omega_m}{16 \pi G_D} \; M_\infty \;
\left(1 - \frac{2 (m - 1)}{m} \; \sum_i a_i \right)
\end{equation}
where $\omega_m$ is the volume of the $m$ dimensional unit
sphere and $G_D$ the $D$ dimensional Newton's constant. The last
term can be ensured to be $< 1$ by choosing $(\sum_i a_i)$ to be
sufficiently small. The expression for $M_{ADM}$ is obtained by
using an effective $m +2$ dimensional metric in Einstein frame,
and also by using the formula given in \cite{jxlu} with a
modification : the formula given there applies to the case where
$\lambda^1 = \cdots = \lambda^n \;$. The term equivalent to $n
\lambda^1$ there is replaced by $\sum_i \lambda^i$ when
$\lambda^i$s are unequal.

\vspace{2ex}

Equations (\ref{freqn}) and (\ref{F(r)}) for $f$ and $e^F$ may
be written in a more convenient form. Define a new variable
$R$ and a constant $b$ by
\begin{equation}\label{Rbdefn} 
R = r^{m - 1} \; \; , \; \; \; \; 
b = \frac{4 (m - 1) \; K}{m} \; \; .
\end{equation}
Then $r (*)_r = (m - 1) \; R (*)_R \;$, where $R-$subscripts
denote derivatives with respect to $R \;$, and equations
(\ref{freqn}) and (\ref{F(r)}) become
\begin{equation}\label{fR} 
b \; \left( 2 (1 - f) - R f_R \right)^2 = 
4 f \; \left( 1 - f - R f_R \right) 
\end{equation}
and 
\begin{equation}\label{e^F}
e^F =  \frac{M_\infty^2 \; f}
{R^2 \; \left( 2 (1 - f) - R f_R \right)^2} \; \; .
\end{equation}
Equation (\ref{bcr}) now means that, in the limit $R \to \infty
\;$,
\begin{equation}\label{bc}
e^F \to 1 
\; \; \; , \; \; \; \; 
f(R) \to 1 - \frac{M_\infty}{R} 
\; \; \; , \; \; \; \; M_\infty = const > 0 \; \; . 
\end{equation}

The quadratic equation (\ref{fR}) can be solved for $R f_R \;$.
The resulting expressions for $R f_R$ and $e^F$ are given, after
a little algebra, by \footnote{ We have taken the solution to
the quadratic equation $\tilde{a} x^2 + \tilde{b} x + \tilde{c}
= 0 \;$ in the form $ x = \frac{- 2 \; \tilde{c}} {\tilde{b} \pm
\sqrt{\tilde{b}^2 - 4 \tilde{a} \tilde{c}}} \;$ which is more
convenient here.}
\begin{equation}\label{RfR}
R f_R = \frac{2 \; (1 - f) \; (f - f_0)}
{f - f_0 \pm \sqrt{\alpha \; f \; (f - f_0)}}
\end{equation}
and
\begin{equation}\label{eF}
e^F = \frac{M_\infty^2 \; 
\left( \sqrt{f - f_0} \pm \sqrt{\alpha \; f} \right)^2}
{4 \; \alpha \; R^2 \; (1 - f)^2} 
\end{equation}
where square roots are always to be taken with a postive sign
and
\begin{equation}\label{alphaetc}
\alpha = \frac{1}{1 + b} \; \; , \; \; \; 
f_0 = 1 - \alpha = \frac{b}{1 + b} \; \; . 
\end{equation}
Among the $\pm$ signs in equation (\ref{RfR}) for $R f_R \;$,
and correspondingly in equation (\ref{eF}) for $e^F \;$, $ \; +$
sign is to be chosen in the limit $R \to \infty \;$ so that, for
any $\alpha > 0 \;$, \footnote{Note that the inequalities on $K$
given in footnote {\bf 3} and the definitions $b = \frac{4 (m -
1) K}{m} \;$ and $\alpha = \frac{1}{1 + b} \;$ imply that $ - \;
\frac{n (m - 1)}{m (n + 1)} \le b \le \infty \;$ and $\frac{m (n
+ 1)}{n + m} \ge \alpha \ge 0 \;$.} one has $R f_R \to 1 - f \;$
in that limit. This is easily checked since $f \to 1 \;$ and $f
- f_0 \to \alpha \;$. This branch choice also gives $R f_R = 1 -
f \;$ in the limit $b \to 0 \;$, equivalently $\alpha \to 1
\;$. This is also easily checked since $f_0 \to 0 \;$ in the
limit $\alpha \to 1 \;$.

Defining a new variable $h$ by $\sqrt{f - f_0} = \epsilon_h
\sqrt{\alpha} \; h $ where $\epsilon_h = Sgn \; h \;$, equation
(\ref{RfR}) becomes
\begin{equation}\label{dRdh}
\frac{d R}{R} = d h \; 
\frac{h \pm \epsilon_h \; \sqrt{1 - \alpha + \alpha h^2}}
{1 - h^2} \; \; , 
\end{equation}
which can be integrated, and thus $R(h)$ obtained, in a closed
form. But this closed form involves $\; ln \;$ and $\; Sinh^{-
1} \;$ terms; it is difficult to invert it to obtain $h(R) \;$;
and its analysis is cumbersome even in special limits. Hence, we
work with equation (\ref{RfR}) itself.


\vspace{4ex}

\centerline{\bf 3. Analysis of solutions}


\vspace{4ex}

\centerline{\bf $K = 0 \;$ case} 

\vspace{2ex}

If there are no compact directions, {\em i.e.} if $n = 0 \;$, or
if $a_i = 0 \;$ for all $i \;$ then we have $2 a_0 = 1 \;$ and
$K = 0 \;$. But $K = 0 \;$ for other choices of $(a_0, a_i) \;$
also, see equation (\ref{AK}). If $K = 0 \;$ then $b = 0 \;$ and
equation (\ref{fR}), together with the boundary conditions
(\ref{bc}), implies that 
\begin{equation}\label{kasner}
1 - f - R f_R = M_R = 0 
\; \; \; \Longrightarrow \; \; \; 
M(R) = M_\infty 
\end{equation}
and, hence, $f = 1 - \frac{M_\infty}{R} \;$. It then follows
from equation (\ref{e^F}) that $e^F = f \;$. Schwarzschild or
black $n-$brane solution follows when $2 a_0 = 1 \;$, and $n = 0
\;$ or $a_i = 0 \;$, but there are solutions for other values of
$(a_0, a_i) \;$ which satisfy $2 A - 1 = K = 0 \;$. Such
solutions, including also the parameter $c$ mentioned in 
footnote {\bf 2}, have been used in \cite{k05} to generate,
following the methods of \cite{t1} -- \cite{t3}, the multi
parameter solutions studied in \cite{m1} -- \cite{m5} in the
context of non BPS branes and tachyon condensation.


\vspace{4ex}

\centerline{\bf $K > 0 \;$ case : $b > 0 \;$, $\; \alpha < 1
\;$}

\vspace{2ex}

From now onwards, we assume that $a_i$ do not all vanish and
that $K \ne 0 \;$, hence $b \ne 0 \;$. Equation (\ref{fR})
implies that $M_R = 1 - f - R f_R \ne 0 \;$ and, hence, the mass
function $M(R) = R \; (1 - f)$ is non trivial. We now study the
solutions $f(R) \;$ to the equation
\begin{equation}\label{RfR+}
R f_R = \frac{2 \; (1 - f) \; (f - f_0)}
{f - f_0 + \sqrt{\alpha \; f \; (f - f_0)}} \; \; . 
\end{equation}
We have chosen the positive square root branch, for reasons
explained below equation (\ref{alphaetc}). Assuming that $f(R)
\to 1 - \frac{M_\infty}{R} \;$ in the limit $R \to \infty \;$,
with $M_\infty > 0$ a constant, we study the behaviour of $f(R)
\;$ as $R$ decreases from $\infty \;$.

For $K < 0 \;$, we are unable to find $f(R)$ for all $R \;$,
with $M_\infty > 0 \;$. Solutions exist with $M_\infty < 0 \;$
which, however, are likely to be of no physical
interest. Therefore, we study only the $K > 0 \;$ case here.

Consider the $K > 0 \;$ case. For $n \ge 2 \;$, $\; K > 0$ can
be ensured by choosing $a_i$ such that $a_i$ do not all vanish
but $\sum_i a_i = 0$. In this case, it follows that $a_0 =
\frac{1} {2} \;$, $\; K = \sum_i a_i^2 > 0 \;$, $\; M_{ADM} >
0 \;$ always, and
\begin{equation}\label{dsnice}
d s^2 = - e^{F} d t^2 + \sum_i e^{2 a_i F} (d x^i)^2
+ \frac{d r^2}{f} + r^2 d \Omega_m^2 \; \; , 
\end{equation}
see equations (\ref{AK}), (\ref{adm}), and (\ref{ds1}). Now $b >
0 \;$ since $K > 0 \;$, and it follows from equation
(\ref{alphaetc}) that $\alpha < 1 \;$ and $f_0 = 1 - \alpha > 0
\;$. We now have from equation (\ref{RfR+}) that $R f_R > 0 \;$
and, hence, $f_R > 0 \;$ for $f_0 < f < 1 \;$.  Therefore, as
$R$ decreases from $\infty \;$, the function $f(R)$ continuously
decreases from $1 \;$.


\vspace{4ex}

\centerline{\bf Evolution of $f(R) \;$ near $R_0$ where $f(R_0)
= f_0 \;$}

\vspace{2ex}

Let $f(R) = f_0 > 0 \;$ at $R = R_0 \;$. As $R$ approaches $R_0
\;$ from above, {\em i.e.} as $R \to R_{0+} \;$, it follows 
from equation (\ref{RfR+}) that 
\begin{equation}\label{fR0+}
R f_R \to 2 \; \sqrt{\frac{\alpha}{1 - \alpha}} \;
\sqrt{f - f_0} \; \to \; 0_+ \; \; . 
\end{equation}
Further, using $R f_{R R} = (R f_R)_R - f_R \;$, and after a
little algebra, it follows that, as $R \to R_{0+} \;$,
\begin{equation}\label{fRR0+}
R f_{R R} \to \frac{2}{R_0} \; \frac{\alpha}{1 - \alpha} 
\; > \; 0 \; \; .
\end{equation}
This implies that, as one goes from $R > R_0 \;$ to $R < R_0
\;$, the derivative $f_R$ goes from positive to negative values,
becoming zero and changing sign at $R_0 \;$. Hence, the function
$f(R)$ decreases for $R > R_0 \;$, reaches a minimum $f_0 > 0
\;$ at $R_0$, and then starts to increase for $R < R_0 \;$.

Now note that the expression inside the square root in equation
(\ref{RfR+}) can be written as
\begin{equation}\label{delta0}
\alpha f (f - f_0) = (f - f_0)^2 + (1 - \alpha) \; (1 - f) \;
(f - f_0) \; > \; (f - f_0)^2 \; \; , 
\end{equation}
the last inequality being valid as long as $f_0 < f < 1 \;$,
which is true near $R_0 \;$ since $1 > f \stackrel{>}{_\sim} f_0
\;$ there. Therefore, one has to choose the negative square root
branch for $R < R_0 \;$ in order to accomodate the change of
sign of $f_R$ at $R_0 \;$. Hence, for $R < R_0 \;$, we have
\begin{equation}\label{RfR-}
R f_R = \frac{2 \; (1 - f) \; (f - f_0)}
{f - f_0 - \sqrt{\alpha \; f \; (f - f_0)}} \; \; . 
\end{equation}
Note that, as $R \to R_{0-} \;$ and $f \to f_0 \;$, the above
equation implies that
\begin{equation}\label{fR0-}
R f_R \to - \; 2 \; \sqrt{\frac{\alpha}{1 - \alpha}} \;
\sqrt{f - f_0} \; \to \; 0_- \; \; . 
\end{equation}

The evolution of $f(R)$ near $R_0$ is similar to that of a
particle trajectory $x(t)$ near a turning point. Let the
particle velocity be $\dot{x} = - \; \sqrt{2 (E - V(x))} \;$, in
an obvious notation. $\; V(x_0) = E \;$ near a turning point
$x_0 \;$ and, as $x \to x_{0+} \;$, $\; E - V(x) \; \propto \;
(x - x_0) \;$ generically. As $x \to x_{0+} \;$, the particle
velocity $\dot{x}$ approaches zero. But its acceleration
$\ddot{x}$ remains finite, non zero, and positive. Hence
$\dot{x}$ changes sign at $x_0$ and becomes $\dot{x} = + \;
\sqrt{2 (E - V(x))} \;$, and the trajectory $x(t)$ reverses its
path.


\vspace{4ex}

\centerline{\bf Evolution of $f(R) \;$ near $R_1$ where 
$R_1 < R_0 \;$ and $f(R_1) = 1 \;$}

\vspace{2ex}

As $R$ decreases below $R_0$, $\; f$ increases above $f_0 $
since $f_R < 0$ for $R < R_0 \;$. Let $f(R_1) = 1 \;$ and $R_1 <
R_0 $. Consider the limit where $R \to R_1 \;$ and $g = 1 - f
\to 0 \;$. Noting that $f - f_0 = \alpha - g \;$ and
\[
\sqrt{\alpha \; f \; (f - f_0)} = \alpha 
- \frac{(1 + \alpha) \; g}{2} + {\cal O}(g^2) \; \; , 
\]
it follows from equation (\ref{RfR-}) that, in the limit $g \to
0 \;$, we have
\begin{equation}\label{fR1}
R f_R = - \frac{4 \alpha}{1 - \alpha} + {\cal O}(g) 
\; < \; 0 \; \; . 
\end{equation}
Note that the above expression is valid for both signs of $g
\;$ in the limit $g \to 0 \;$; equivalently for both $f < 1 \;$
and $f > 1 \;$ cases in the limit $f \to 1 \;$. Thus, $f_R(R_1)
\;$ remains negative and non zero which implies that as $R$
approaches $R_1$ and decreases further, the function $f$
approaches $1$ and increases further.

In equation (\ref{RfR-}) for $R f_R \;$, the numerator is
positive for $f < 1 \;$ and negative for $f > 1 \;$. However,
the denominator which is negative for $f < 1 \;$ becomes
positive for $f > 1 \;$ since $\alpha f (f - f_0) \; < \; (f -
f_0)^2 \;$ for $f > 1 \;$, see equation (\ref{delta0}). Hence,
$R f_R \;$ is negative for both $f < 1$ and $f > 1 \;$. This is
also clear from equation (\ref{fR1}) since it is valid for both
signs of $g \;$ in the limit $g \to 0 \;$. In particular, it
follows that $R f_R < 0 \;$ and $f > 1 \;$ for $R < R_1 \;$.


\vspace{4ex}

\centerline{\bf Evolution of $f(R) \;$ in the limit $f \to
\infty \;$}

\vspace{2ex}

As $R$ decreases below $R_1$, $\; f$ increases above $1 \;$.
Consider the limit $f \gg 1 \;$. It then follows from equation
(\ref{RfR-}) that
\begin{equation}\label{fR>>1}
R f_R \simeq - \; \frac{2 \; f}{1 - \sqrt{\alpha}}
\; \; \; \Longrightarrow \; \; \; 
f \simeq (const) \; \left( \frac{M_\infty}{R} 
\right)^{\frac{2}{1 - \sqrt{\alpha}}} 
\end{equation}
and, hence, that $f \to \infty \;$ as $R \to 0 \;$. 

\vspace{2ex}

To summarise: the evolution of $f(R)$ for $b > 0 \;$ is as
follows. As $R$ decreases from $\infty$ to $0 \;$, $ \; f(R)$
decreases from $1 \;$, reaches a minimum $f_0 = 1 - \alpha > 0
\;$ at $R = R_0 \;$, increases to $1$ again at $R = R_1 < R_0
\;$, and then increases to $\infty$ as $f \sim R^{- \;
\frac{2}{1 - \sqrt{\alpha}}} \;$ in the limit $R \to 0 \;$.


\vspace{4ex}

\centerline{\bf Evolution of $e^F \;$} 

\vspace{2ex}

The evolution of $e^F$ can be easily read off from equations
(\ref{e^F}) and (\ref{eF}), which we reproduce below:
\begin{eqnarray}
e^F & = & \frac{M_\infty^2 \; f} 
{R^2 \; \left( 2 (1 - f) - R f_R \right)^2} 
\;\; \; \; \; 
\; \; \; for \; \; \; 0 < R < \infty  \label{e^F1} \\
& = & \frac{M_\infty^2 \; 
\left( \sqrt{f - f_0} + \sqrt{\alpha \; f} \right)^2}
{4 \; \alpha \; R^2 \; (1 - f)^2} 
\; \; \; for \; \; \; R_0 < R < \infty \label{eF+} \\
& = & \frac{M_\infty^2 \; 
\left( \sqrt{f - f_0} - \sqrt{\alpha \; f} \right)^2}
{4 \; \alpha \; R^2 \; (1 - f)^2} 
\; \; \; for \; \; \; 0 < R < R_0   \; \; .   \label{eF-} 
\end{eqnarray}
The behaviour of $e^F$ in the limit $R \to \infty \;$ is given
by equation (\ref{asympeF}). It can be checked that $e^F$
remains non zero and finite for $0 < R < \infty \;$, in
particular at $R_0$ and $R_1 \;$; and that, in the limit $R \to
0 \;$ where $f \gg 1 \;$,
\begin{equation}\label{e^F0}
e^F \; \sim \; \frac{M_\infty^2}{R^2 \; f} 
\sim \; \left( \frac{M_\infty}{R} 
\right)^{- \frac{2 \sqrt{\alpha}} {1 - \sqrt{\alpha}}}
\; \; \to \; \; 0 \; \; . 
\end{equation}

It can be shown that $F_R \ne 0 \;$ for $R < \infty \;$. If $F_R
= 0$ then it follows from equation (\ref{aa1}), and then from
equation (\ref{fR}), that
\[
2 (1 - f) - R f_R = 0 = 1 - f - R f_R
\; \; \; \Longrightarrow \; \; \; 
1 - f = R f_R = 0 \; \; . 
\]
This is the case at $R = \infty \;$. From the evolution of $f(R)
\;$, we have $f_R = 0$ but $1 - f = 1 - f_0 = \alpha \ne 0 \;$
at $R = R_0 \;$, and $1 - f = 0 \;$ but $R f_R \ne 0 \;$ at $R =
R_1 \;$. Thus, besides at $R = \infty \;$, we see from the
evolution of $f(R)$ that $1 - f$ and $R f_R$ do not both vanish
and, hence, that $F_R \;$ cannot vanish. The asymptotic
behaviour of $e^F$ given in equations (\ref{asympeF}) and
(\ref{e^F0}) in the limits $R \to \infty \;$ and $R \to 0 \;$
then implies that $e^F$ decreases monotonically from $1$ to $0$
as $R$ decreases from $\infty$ to $0 \;$.

It can further be shown that $e^F$ always remains $< f \;$. Note
from equations (\ref{asympeF}) that $e^F < f \;$ in the limit $R
\to \infty \;$. It also follows, using equations (\ref{aa1}) and
(\ref{RfR+}), that $f \; (R F_R - R f_R) > 0 \;$, and hence $F_R
> f_R \;$, for $R_0 < R < \infty \;$ where $f_0 < f < 1 \;$. It
then follows that $e^F < f $ for $R_0 < R < \infty \;$. Since
$e^F$ continues to decrease and $f$ increases above $f_0$ for $R
< R_0 \;$, it follows that $ e^F < f$ for all $R \;$.


\vspace{4ex}

\centerline{\bf Evolution of the mass function $M(R) \;$} 

\vspace{2ex}

The mass function is defined by $M(R) = R \; (1 - f) \;$. Since
$M_R = 1 - f - R f_R \;$, it follows from equation (\ref{fR})
that $M_R = 0$ and $M(R)$ is constant if and only if $b = 0
\;$. For $b > 0 \;$, it follows from the evolution of $f(R) \;$
that $M(R)$ is a positive constant $= M_\infty \;$ at $R =
\infty \;$, remains positive for $R_1 < R < \infty \;$, vanishes
at $R = R_1 \;$, becomes negative for $R < R_1 \;$, and, in the
limit $R \to 0 \;$ where $f \gg 1 \;$,
\[
M(R) \; \sim - \; R f \; \sim - \; 
R^{- \frac{1 + \sqrt{\alpha}} {1 - \sqrt{\alpha}}}
\; \; \to \; \; - \; \infty \; \; . 
\]

If $\; b > 0 \;$ then it follows, for the same reasons as in the
case of $F_R \;$, that $M_R = 1 - f - R f_R$ cannot vanish for
$R < \infty \;$. Its asymptotic behaviour described above then
implies that $M(R)$ decreases monotonically from a positive
constant $M_\infty$ to $- \; \infty \;$ as $R$ decreases from
$\infty$ to $0 \;$. We point out here that solutions with
negative masses in the interior also occur in \cite{zurek,
thooft} which study the back reaction of Hawking radiation in
four dimensional spacetime; and in \cite{date1} which study
static solutions with incoming radiation matching the outgoing
one if such solutions are matched onto negative mass
Schwarzschild ones.


\vspace{4ex}

\centerline{\bf Summary of the solutions}

\vspace{2ex}

In summary, we have the $D = n + m + 2$ dimensional metric
components, with $m \ge 2 \;$ and corresponding to static
brane--like solutions, given by
\[
- g_{t t} = e^{2 a_0 F}  
\; \; , \; \; \; 
g_{i i} = e^{2 a_i F}  
\; \; , \; \; \; 
g_{r r} = \frac{1}{f} 
\]
which are all functions of $R = r^{m - 1} \;$, with $r$ denoting
the physical size of the $m$ sphere. The solutions are all
required to have positive ADM mass and the asymptotic behaviour
given in equation (\ref{bcr}) in the limit $r \to \infty \;$. We
also have
\[
A = a_0 + \sum_i a_i = \frac{1}{2} \; \; , \; \; \; 
K = a_0^2 + \sum_i a_i^2 - \frac{1}{4} 
\]
and the definitions $b = \frac{4 (m - 1) K}{m} \;$ and $\alpha =
\frac{1}{1 + b} \;$. The standard Schwarzschild solution follows
for $2 a_0 - 1 = a_i = 0 \;$. For other values of $(a_0, a_i)
\;$ but with $b = 0 \;$, there exist more general solutions. In
all these solutions, the metric components vanish or diverge at
a non zero, finite value of $R = R_h \; $, which is either a
regular horizon or, possibly, a curvature singularity depending
on the values of $(a_0, a_i) \;$. In all these solutions, the
mass function $M(R) = R \; (1 - f) \;$ remains constant.

We assume that $b \ne 0 \;$ generically. Then $M(R)$ is non
trivial and cannot remain constant. Further assuming that $b > 0
\;$, we have described the evolution of $f$, $\; e^F$, and $M
\;$. Note that $f$ and $e^F$ and, hence, all metric components
remain non zero and finite for $0 < R \le \infty \;$. This
implies that there is no horizon, and that the curvature
invariants are all finite, for $0 < R \le \infty \;$. 

Consider the limit $R \to 0 \;$. In this limit, we have $f \gg 1
\;$ and $e^F \to 0 \;$. Consider the Riemann tensor components
$R_{a b c d} = e^M_a e^N_b e^P_c e^Q_d \; R_{M N P Q}$ in local
tangent frame coordinates. For the $D$ dimensional metric given
by equation (\ref{ds}), the non vanishing components of $R_{a b
c d}$ are given by
\begin{eqnarray}
R_{ r i' r j'} & = & - \delta_{i' j'} \; \; 
e^{- 2 \lambda} \left( \lambda_{r r}^{i'} 
+ (\lambda_r^{i'} - \lambda_r) \; \lambda_r^{i'} \right)
\label{rirj} \\
R_{ r a r b} & = & - h_{a b} \; \; e^{- 2 \lambda} \left(
\sigma_{r r} + (\sigma_r - \lambda_r) \; \sigma_r \right)
\label{rarb} \\
R_{i' j' k' l'} & = & (\delta_{i' l'} \delta_{j' k'} -
\delta_{i' k'} \delta_{j' l'}) \; \;
e^{- 2 \lambda} \left( \lambda_r^{i'} \lambda_r^{j'} \right)
\label{ijkl} \\
R_{i' a j' b} & = & - \delta_{i' j'} h_{a b} \; \; 
e^{- 2 \lambda} \lambda_r^{i'} \sigma_r \label{iajb} \\
R_{a b c d} & = & e^{- 2 \sigma} \rho_{a b c d}(h) 
+ (h_{a d} h_{b c} - h_{a c} h_{b d}) \; \; 
e^{- 2 \lambda} \sigma_r^2 \label{abcd}
\end{eqnarray}
where $i' = (0, i) \;$, $\lambda^{i'} = (\psi, \lambda^i) \;$,
$h_{a b} \;$ is the metric on the $m$ dimensional unit sphere
given by $d \Omega_m^2 = h_{a b} d \theta^a \theta^b \;$, and
$\rho_{a b c d}(h)$ is the corresponding Riemann tensor. It can
now be seen that, in the limit $R \to 0 \;$, we have
\begin{equation}\label{rabcd0}
R_{a b c d} \; \sim \; \frac{f}{r^2} 
\; \; \to \; \infty \; \; .  
\end{equation}
This implies that the tidal forces diverge and there is a
curvature singularity in the limit $R \to 0 \;$.

Note that the presence and the role of the $n-$dimensional space
is crucial for these properties of the solutions. The absence of
the $n-$dimensional space, or the trivialty of its metric, means
that $a_i = b = 0 \;$, thus leading to the standard
Schwarzschild or black $n-$brane solution. We have assumed that
$a_i \ne 0 \;$ generically and, further, that $b > 0 \;$ which
then lead to the present solutions.


\vspace{4ex}

\centerline{\bf 4. Physical relevance of the solutions}

\vspace{2ex}

Physical relevance of the present solutions can be naturally
motivated and, indeed, such solutions may be naturally
anticipated if one assumes that Mathur's fuzzball proposal for
black holes is correct. See \cite{f1} -- \cite{f5} for a review
of this proposal. Broadly speaking, according to this proposal,
the black hole entropy arises due to the microstates of M theory
objects, equivalently string theory objects, which are typically
bound states of intersecting brane configurations with a large
number of low energy excitations living on them. For example, an
effective four dimensional black hole may be described by a
$22'55'$ configuration which consists of two sets of $M2$ branes
and two sets of $M5$ branes, intersecting according to BPS
rules.

According to the fuzzball picture, the spacetime described by
such brane configurations is indistinguishable from that of
black holes at large distances, typically larger than ${\cal
O}(1) \;$ times the Schwarzschild radius. At shorter distances,
the spacetime is different from that of black holes and, in
particular, has no horizon.

If this picture is correct then it should be possible to
construct a star, modelling its M theory brane constituents by
appropriate matter sources. At large distances, it should appear
as a spherically symmetric four dimensional (more generally, $(m
+ 2)$ dimensional) star; should have a finite radius, be stable,
and have no horizon irrespective of how high its mass $M_*$ is;
and the thermodynamics of its constituents should give an
entropy $\propto M_*^2 \;$.

Technically, one constructs the interior of the star and, at its
surface, matches the interior solution onto vacuum solutions. If
the matching vacuum solution is the standard Schwarzschild one
then, for any choice of matter sources that the author can think
of, it seems impossible to obtain a star solution with the above
properties. Also, such a matching seems to miss a crucial
ingredient : that, at a fundamental level, both the spacetime
and the constitutents of the star are higher dimensional and
this higher dimensionality is likely to play an important role.

In \cite{k07, bdr, br}, we had studied early universe using
$22'55'$ intersecting brane configuration. Starting with a
eleven dimensional universe, we found that, at later times, the
seven toroidal brane directions cease to expand or contract and
stabilse to constant sizes; and, in the limit $t \to \infty \;$,
the corresponding metric components $e^{\lambda^i} \to e^{v^i}
\; (1 + \frac{c(t)} {t^\delta} ) \;$ where $v^i$ and $\delta >
0$ are constants and $\vert c(t) \vert \;$ is finite. This
results in an effectively four dimensional expanding universe.
The tailing--off behaviour of $e^{\lambda^i}$ suggests that, in
the context of stars also, the internal directions are likely to
have non trivial $r$ dependence in the limit $r \to \infty \;$.

This line of reasoning is what led us to study the higher
dimensional vacuum solutions, in particular to study the general
solutions with non trivial dependence of $e^{\lambda^i} \;$. It
turned out that such solutions exist indeed, with the properties
described in this paper. The early universe study mentioned
above also suggests that stars whose exterior solutions are
similar to the ones presented here may form in a physical
collapse, and that one has to carefully take into account the
higher dimensional nature of the constituents.


\vspace{4ex}

\centerline{\bf 5. Conclusions}

\vspace{2ex}

Finding the more general vacuum solutions is only a
beginning. It is important to actually construct both
equilibrium and collapsing star solutions, study their
stability, thermodynamic entropy, and other properties.

Also, it will be interesting to generalise the present solutions
to include rotation and charges. One may also start from the
present solutions and, using the techniques of {\em e.g.}
\cite{t1, t2, t3, k05}, generate string and M theory brane
solutions.

\vspace{4ex}

{\bf Acknowledgement:} 
We thank Steven G. Avery for discussions and for pointing out
references \cite{zurek, thooft}.


\end{document}